\documentclass[aps,prl,showpacs,superscriptaddress,twocolumn]{revtex4}

\usepackage{graphicx}
\usepackage{longtable}
\usepackage[ansinew]{inputenc}

\begin{document}

\title{Controlling the Kondo Effect in CoCu$_{\mathbf{n}}$ Clusters Atom by Atom}

\author{N. N\'{e}el}
\author{J. Kr\"{o}ger}\email{kroeger@physik.uni-kiel.de}
\author{R. Berndt}
\affiliation{Institut f\"{u}r Experimentelle und Angewandte Physik,
Christian-Albrechts-Universit\"{a}t zu Kiel, D-24098 Kiel, Germany}
\author{T. Wehling}
\author{A. Lichtenstein}
\affiliation{Institut f\"{u}r Theoretische Physik I,
Universit\"{a}t Hamburg, D-20355 Hamburg, Germany}
\author{M. I. Katsnelson}
\affiliation{Institute for Molecules and Materials,
Radboud University Nijmegen, NL-6525 AJ Nijmegen, The Netherlands}

\begin{abstract}
Clusters containing a single magnetic impurity were investigated by scanning
tunneling microscopy, spectroscopy, and {\it ab initio} electronic structure
calculations. The Kondo temperature of a Co atom embedded in Cu clusters on
Cu(111) exhibits a non-monotonic variation with the cluster size. Calculations
model the experimental observations and demonstrate the importance of the local
and anisotropic electronic structure for correlation effects in small clusters.
\end{abstract}

\date{\today}


\maketitle

Nanometer scaled electronic devices require the understanding and the control
of electron behavior, in particular of correlation effects, at the atomic scale.
Experimental techniques such as scanning tunneling microscopy (STM) and
spectroscopy and angle-resolved photoemission demonstrate the relevance of
many-body phenomena beyond standard band theory \cite{epl_02}. In many cases,
especially for compounds of $d$ and $f$ elements, strong electron correlations
should be taken into account even for a qualitative description of electron
energy spectra and physical properties \cite{gko_06}. The Kondo effect, which
is related to the resonant scattering of conduction electrons by quantum local
centers, is one of the key correlation effects in condensed matter physics
\cite{ahe_93}. It arises from the interaction of a single localized
magnetic moment with a continuum of conduction electrons and results in a
sharp Abrikosov-Suhl resonance of the local impurity spectral function at
the Fermi level and below a characteristic Kondo temperature,
$T_{\text{K}}$ \cite{ahe_93}. Originally detected as a resistance increase
below $T_{\text{K}}$ of materials with dilute magnetic impurities, this
many-body phenomenon has been observed for semiconductor quantum dots
\cite{dgo_98,scr_98} and for atoms \cite{jli_98,vma_98,hcm_00,nkn_02} and
molecules \cite{azh_05} at surfaces. Whereas the understanding of
correlation phenomena of bulk materials and single impurities has made
significant progress much less is known about these effects in structures
such as clusters where typical dimensions are on an atomic scale.  STM
offers unique opportunities to fabricate such structures by manipulation of
single atoms and a few attempts have been made to achieve a degree of
control over the electronic \cite{jkl_00,nni_02,jla_07} and magnetic
structure \cite{tja_01,ahe_04} of atoms and clusters on surfaces.

Here we show that the Kondo effect of a magnetic impurity does not simply
scale with the number of nearest neighbors. Rather, the local and anisotropic
electronic structure at the impurity site is important for correlations in
atomic scale structures, where ``each atom counts'' \cite{uhe_99}. We investigate
the evolution of the Abrikosov-Suhl resonance induced by a Co atom adsorbed
on Cu(111) as we change the number of nearest neighbor Cu atoms. A non-monotonic
behavior of the Kondo temperature with the cluster size is at variance with
bulk models which are typically invoked to interpret the Kondo effect of a
single adsorbed atom (adatom).  Scanning tunneling spectroscopy
measurements and {\it ab initio} calculations clearly reveal that the Kondo
effect in atomic scaled clusters depends crucially on their detailed geometric
structure and thus on the local density of conduction electron states at the
magnetic site.

Experiments were performed with a home-made scanning tunneling microscope
operated at $7\,\text{K}$ and in ultra-high vacuum with a base pressure of
$10^{-9}\,\text{Pa}$. Tungsten tips and Cu(111) surfaces were prepared by
argon ion bombardment and annealing. While single Co atoms were deposited onto
the cold surface using an electron beam evaporator and an evaporant of $99.99\,\%$
purity, single Cu atoms were transferred from the tip as previously reported
\cite{lli_05}. The individual atoms were chemically identified by the presence
(Co) or absence (Cu) of the Abrikosov-Suhl resonance. Clusters consisiting of
a single Co atom and several Cu atoms \cite{jla_07,jst_06} were fabricated by
atom manipulation with the microscope tip. Spectroscopy was performed by a
state-of-the-art lock-in technique.
\begin{figure}
  \includegraphics[width=85mm]{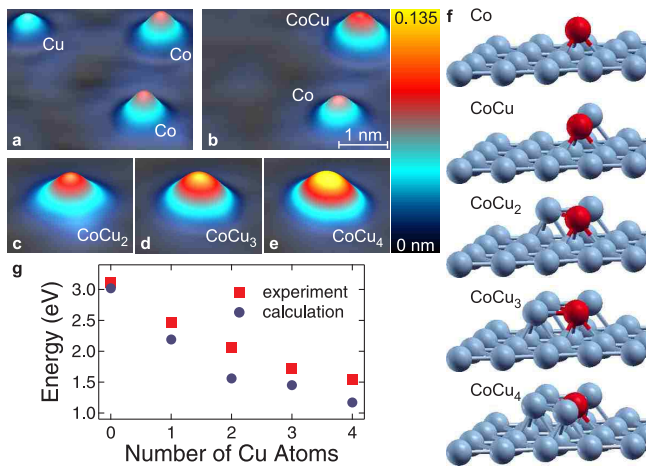}
  \caption{(Color online) Scanning tunneling microscopy images and resonance
  energies of CoCu$_n$ clusters acquired at $7\,\text{K}$. (a) STM images of
  single Co and Cu atoms adsorbed on Cu(111) prior to fabricating CoCu$_n$
  clusters. (b) - (e) Sequence of STM images showing CoCu$_n$ ($n=1,\ldots,4$)
  clusters. The same lateral and height scale was used for each image. Images
  were acquired at a sample voltage of $100\,\text{mV}$ and a tunneling current
  of $0.1\,\text{nA}$. (f) Stick-and-ball models of fully relaxed adsorption
  structures of CoCu$_n$ ($n=0,\ldots,4$) on Cu(111). (g) Experimental energy
  of unoccupied cluster state versus the cluster size. The size of the symbols
  (squares) corresponds to the uncertainty of the energies. Calculated energies
  are depicted as circles.}
  \label{fig1}
\end{figure}

Figures \ref{fig1}(a)-(e) show a series of constant-current STM images, which
illustrates fabrication of a CoCu cluster [Figs.\,\ref{fig1}(a), (b)] and
dimensions of CoCu$_n$ ($n=1,\ldots,4$) clusters [Figs.\,\ref{fig1}(b)-(e)].
In STM images acquired at a sample voltage of $100\,\text{mV}$ Co atoms appear
higher than Cu atoms [Fig.\,\ref{fig1}(a)], and thus an additional means of
discriminating the adatom species is provided. The stick-and-ball models of
the clusters used for the calculations are presented in Fig.\,\ref{fig1}(f).
Spectroscopy of the differential conductance ($\text{d}I/\text{d}V$) performed
with the tip positioned above the cluster center reveals an unoccupied state
whose energy decreases with increasing number of Cu atoms [squares in
Fig.\,\ref{fig1}(g)]. Our calculations show that this resonance is of $p_z$
character and are in agreement with experimental data [circles in
Fig.\,\ref{fig1}(g)]. This resonance likewise serves as an indicator of the
cluster composition and size.

\begin{figure}
  \includegraphics[width=80mm]{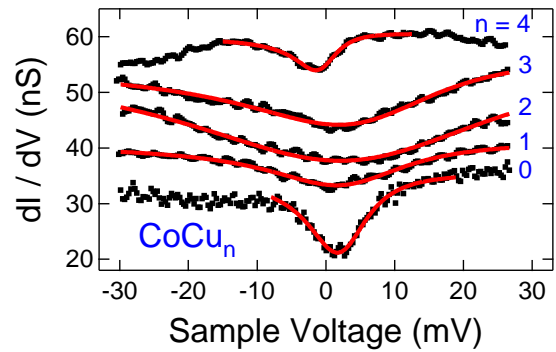}
  \caption{(Color online) Spectroscopy of the differential conductance
  ($\text{d}I/\text{d}V$) of CoCu$_n$ clusters around the Fermi level performed
  at $7\,\text{K}$. The spectroscopic feature is the Abrikosov-Suhl resonance
  induced by the Kondo effect. Solid lines depict fits of a Fano line shape
  (Eq.\,\ref{fano}) to experimental data. Prior to spectroscopy the tunneling
  gap was set at $30\,\text{mV}$ and $1\,\text{nA}$. Spectra for $n=1,2,3,4$
  are vertically offset by $5$, $10$, $20$, $25\,\text{nS}$, respectively.}
  \label{fig2}
\end{figure}

Figure \ref{fig2} shows a sequence of $\text{d}I/\text{d}V$ spectra
acquired on a single Co atom (lower curve, $n=0$) and above the center of
CoCu$_n$ clusters with $n$ ranging between $1$ and $4$. Starting from a
sharp indentation of the differential conductance close to the Fermi level
(sample voltage $V=0\,\text{mV}$) which is the spectroscopic signature of
the single-Co Kondo effect on Cu(111), the $\text{d}I/\text{d}V$ curve
broadens appreciably upon adding two Cu atoms to the Co atom. Surprisingly,
upon increasing the number of hybridizations to three and four Cu atoms,
$\text{d}I/\text{d}V$ spectra exhibit a sharpening of the resonance again
in contrast with the monotonic behaviour of the unoccupied $p_z$-like state
with the number of copper atoms [Fig.\,\ref{fig1}(g)].

\begin{figure}
  \includegraphics[width=80mm]{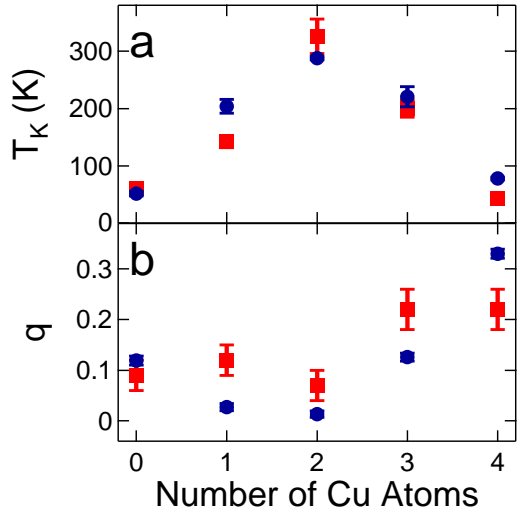}
  \caption{(Color online) Experimental (squares) and calculated (circles)
  Kondo temperatures and asymmetry parameters. (a) Kondo temperature
  ($T_{\text{K}}$) as a function of cluster size. (b) Asymmetry parameter
  ($q$) as a function of cluster size.}
  \label{fig3}
\end{figure}

To quantify the broadening of the resonances at the Fermi level we described
experimental data by a Fano line shape \cite{ufa_61}:
\begin{equation}
  \frac{\text{d}I}{\text{d}V}\propto\frac{(q+\epsilon)^2}{1+\epsilon^2}
  \label{fano}
\end{equation}
with $q$ the asymmetry parameter of the Fano theory and
$\epsilon=(\text{e}V-\epsilon_{\text{K}})/(\text{k}_{\text{B}}T_{\text{K}})$
where $-\text{e}$ denotes the electron charge, $V$ the sample voltage,
$\epsilon_{\text{K}}$ the resonance energy, and $\text{k}_{\text{B}}$
Boltzmann's constant. The profiles of Fano line shapes according to
Eq.\,\ref{fano} and fitted to $\text{d}I/\text{d}V$ data are presented as
solid lines in Fig.\,\ref{fig2}. The striking behavior of broadening and
sharpening of the Abrikosov-Suhl resonance with increasing number of Cu atoms
is reflected by the Kondo temperature $T_{\text{K}}$ which we plotted in
Fig.\,\ref{fig3}(a). While the single Co adatom exhibits
$T_{\text{K}}=(61\pm 4)\,\text{K}$, which is in agreement with a previous
publication \cite{nkn_02}, for CoCu$_2$ we find $(326\pm 30)\,\text{K}$ which
then decreases again to $(43\pm 6)\,\text{K}$ for CoCu$_4$. Figure \ref{fig3}(b)
shows the evolution of the asymmetry parameter with the cluster size. For
$n=0,1,2$ the asymmetry parameter varies weakly around $0.1$. It increases
steeply by $\approx 100\,\%$ upon adding a third Cu atom. Kondo temperatures
and asymmetry parameters for the various clusters are summarized in Table
\ref{tab1}.
\begin{table}
  \caption{Experimental and calculated Kondo temperatures ($T_{\text{K}}$)
  and asymmetry parameters ($q$) for clusters CoCu$_n$.
  $T_{\text{K}}$ and $q$ were extracted from fits of the Fano line shape
  (Eq.\,\ref{fano}) to measured Kondo resonances shown in Fig.\,\ref{fig2}.
  Experimental uncertainty margins correspond to standard deviations obtained
  by a statistical analysis of a variety of fits. Calculated $T_{\text{K}}$
  and $q$ are the arithmetic mean of values obtained by a Gaussian broadening
  of the $sp$ local density by $100\,\text{meV}$ and $50\,\text{meV}$. The
  uncertainty margins reflect the deviations of the arithmetic mean from
  data obtained by $100\,\text{meV}$ and $50\,\text{meV}$ broadening.}
  \begin{ruledtabular}
    \begin{tabular}{ccccccc}
          & \multicolumn{2}{c}{Experiment}      & \multicolumn{2}{c}{Calculation}     \\
      $n$ & $T_{\text{K}}$ (K) & $q$            & $T_{\text{K}}$ (K) & $q$            \\
      0   & $61\pm 4$          & $0.09\pm 0.03$ & $52\pm 4$          & $0.12\pm 0.01$ \\
      1   & $142\pm 10$        & $0.12\pm 0.03$ & $204\pm 12$        & $0.03\pm 0.01$ \\
      2   & $326\pm 30$        & $0.07\pm 0.03$ & $288\pm 3$         & $0.01\pm 0.01$ \\
      3   & $200\pm 15$        & $0.22\pm 0.04$ & $221\pm 18$        & $0.13\pm 0.01$ \\
      4   & $43\pm 6$          & $0.22\pm 0.04$ & $79\pm 3$          & $0.33\pm 0.01$
    \end{tabular}
  \end{ruledtabular}
  \label{tab1}
\end{table}

The peculiar behavior of the Kondo resonance with cluster size is at variance
with the monotonic dependence of $T_{\text{K}}$ on the average hybridization
strength expected from bulk models of the Kondo effect. Theoretical approaches
to the Kondo effect of magnetic impurities on surfaces consider the importance
of bulk \cite{ouj_00,cli_06} and surface \cite{gfi_03,oag_01,dpo_01} states
in the scattering of electron waves at the magnetic impurity site.
Recent experiments with Co adatoms on Cu(100) \cite{nkn_02}, which does not
host any surface state close to the Fermi level \cite{cba_03}, or on Cu(111)
but in the vicinity of atomic surface steps that affect the surface states
\cite{lli_04} have indicated the importance of bulk rather than surface states
for the Kondo effect. In this situation, {\it ab initio} electronic structure
calculations which are not restricted by specific model assumptions can be
very useful. Here, we perform accurate calculations of the local electronic
structure of clusters on the copper surface.

The Kondo temperature for a single magnetic impurity can be estimated as
\begin{equation}
  T_{\text{K}}\approx W\,\sqrt{|J\,N(E_{\text{F}})|}\,\exp\left(-\frac{1}{|J\,N(E_{\text{F}})|}\right)
  \label{TK}
\end{equation}
where $W$ is the conduction ($s$) electron bandwidth, $J$ the $s$-$d$
exchange integral, and $N(E_{\text{F}})$ the density of states at the Fermi
energy $E_{\text{F}}$ \cite{ahe_93}. Since the $s$-$d$ exchange interaction is
local, $N(E_{\text{F}})$ in Eq.\,\ref{TK} is the local density of states of
conduction electrons at the magnetic atom.

The asymmetry parameter, $q$, is also determined by the local electronic
structure. In a simple model $q$ may be expressed as
\begin{equation}
  q = \frac{\gamma + \text{Re}\,G(E_{\text{F}})}{\text{Im}\,G(E_{\text{F}})},
  \label{qF}
\end{equation}
where $G$ is the local Green's function of the conduction electrons at the
impurity site and $\gamma$ measures the ratio of the coupling of the
scanning tunneling microscope tip to conduction electron states and to the
strongly localized Co $d$ states \cite{vma_01}.
\begin{figure}
  \includegraphics[width=80mm]{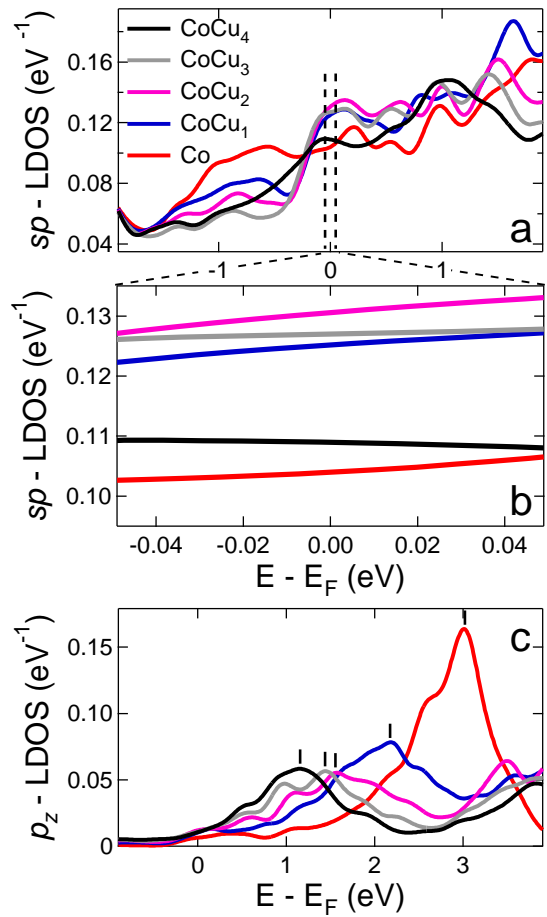}
  \caption{(Color online) Calculated local density of states (LDOS) at the Co
  site for CoCu$_n$ assemblies ($n=0,\ldots,4$). (a) $sp$ LDOS plotted for
  energies between $-1.9\,\text{eV}$ and $1.9\,\text{eV}$ with respect to the
  Fermi energy. (b) Close-up view of (a) showing the energy scale of the Kondo
  resonance. (c) $p_z$ LDOS with indicated energies of unoccupied cluster
  resonances shown in Fig.\,\ref{fig1}(g).}
  \label{fig4}
\end{figure}

To realistically describe the CoCu$_n$ clusters on Cu(111)
we calculated their electronic and structural properties by means of density
functional theory using the generalized gradient approximation to the exchange
correlation potential \cite{jpe_92}. For solving the resulting Kohn-Sham
equations accurately, the Vienna Ab Initio Simulation Package \cite{gkr_94}
with the projector augmented wave basis sets \cite{gkr_99,pbl_94} and
$350\,\text{eV}$ as plane wave cut-off have been used. We modeled the CoCu$_n$
structures using $4\times 3$ supercells of Cu(111) slabs containing up to $7$
Cu layers. These structures haven been fully relaxed with the requirement
that all forces are less than $0.2\,\text{eV}\,\text{nm}^{-1}$ and were then
used to calculate the local density of states. The supercell Brillouin zone
integrations for obtaining the local density of states have been performed
using the tetrahedron-method on $6\times 6$ $k$ meshes with subsequent
$50$ - $100\,\text{meV}$ Gaussian broadening. Below, we use both, $50$ and
$100\,\text{meV}$ smearing, to provide an uncertainty margin for the derived
Kondo temperatures and asymmetry parameters.

The relaxed structures [Fig.\,\ref{fig1}(f)] show that the distance of the Co
atom to the Cu surface increases monotonically upon adding Cu atoms to
the cluster. However, the local electronic structure at the Fermi level and
at the Co site varies in a non-monotonic way with increasing number of Cu
atoms. The substrate conduction band states extend to the Co adatom site
where they interact with localized Co $3d$ electrons. This interaction is
quantified by the Co $sp$-projected local density of states, $N(E)$, which
is shown in Figs.\,\ref{fig4}(a) and \ref{fig4}(b). The dependence of
$G(E_{\text{F}})$ on the number of Cu atoms is non-monotonic [see the close-up
view around $E_{\text{F}}$ in Fig.\,\ref{fig4}(b)] and leads, according to
Eq.\,\ref{TK}, to the experimentally observed non-monotonic behavior of the
Kondo temperature (Fig.\,\ref{fig3}). We used $W=20\,\text{eV}$ and
$J=1.3\,\text{eV}$ as fitting parameters. Moreover, with
$\gamma=0.22\,\text{eV}^{-1}$ in Eq.\,\ref{qF} the variation of $G(E_{\text{F}})$
with the number Cu atoms also reproduces the non-monotonic trend of $q$
observed in the experiments [Fig.\,\ref{fig3}(b)]. Figure \ref{fig4}(c) shows
the evolution of unoccupied resonance energies with the number of Cu atoms.
The resonances appear as peaks in the calculated $p_z$-projected local density
of states whose energies show the same monotonic behavior as in the experiment
[Fig.\,\ref{fig1}(g)].

The non-monotonic dependence of $N(E_{\text{F}})$ on the number of Cu atoms
surrounding the Co atom is related to the specific nature of chemical bonds
in the system. Since localized $d$ orbitals of Co atoms are strongly anisotropic
and form well-defined directional bonds the degree of their hybridization with
Cu atoms is determined by the whole geometry of the cluster rather than by
just the coordination number.

In summary, we have shown that the Kondo effect in clusters is
but depends crucially on their detailed geometric structure. Atom by atom
manipulation changes the local density of conduction electron states at the
magnetic site and thus varies the Kondo temperature in rather broad limits.
The results demonstrate that electron correlations may be tuned in atomic-scale
structures.

Discussion with H.\ Kroha (University of Bonn, Germany) and financial support
by the Deutsche Forschungsgemeinschaft through SFB 668 and SPP 1153 are
gratefully acknowledged.

\end{document}